\documentclass[final,5p,times,twocolumn]{elsarticle}

\usepackage{lineno,hyperref}
\modulolinenumbers[1]
\journal{Pattern Recognition Letters}

\usepackage[utf8]{inputenc} 
\usepackage[T1]{fontenc}    
\usepackage[]{hyperref}       

\usepackage{booktabs}       
\usepackage{amsfonts}       
\usepackage{nicefrac}       
\usepackage{microtype}      
\usepackage{xcolor}         

\usepackage{graphicx}
\usepackage{subfigure}
\usepackage{multirow}
\usepackage[export]{adjustbox}
\usepackage{enumerate}
\usepackage{amsmath,amssymb,amsthm,mathtools}
\usepackage{caption}
\usepackage{dsfont}
\usepackage{textcomp}
\usepackage[multiple]{footmisc}
\usepackage{algorithm,algorithmic}
\usepackage{wrapfig}
\usepackage{enumitem}
\usepackage{grffile}
\usepackage{threeparttable}
\usepackage{pifont}
\usepackage{bm}

\theoremstyle{definition}
\newtheorem{definition}{Definition}
\newtheorem*{definition*}{Definition}

\newtheorem*{example*}{Example}
\theoremstyle{theorem}
\newtheorem{theorem}[definition]{Theorem}

\newtheorem*{theorem*}{Theorem}
\newtheorem*{lemma*}{Lemma}
\newtheorem*{proposition*}{Proposition}
\newtheorem*{corollary*}{Corollary}
\newtheorem*{remark*}{Remark}
\newtheorem*{claim*}{Claim}
\newtheorem*{problem*}{Problem}
\newtheorem*{observation*}{Observation}

\DeclareMathOperator*{\argmin}{arg\,min}

\allowdisplaybreaks

\begin{document}

\begin{frontmatter}

\title{Sequential Likelihood-Free Inference with Neural Proposal}

\author[add1]{Dongjun Kim}
\author[add2]{Kyungwoo Song}
\author[add1]{Yoon-Yeong Kim}
\author[add3]{Yongjin Shin}
\author[add4]{Wanmo Kang}
\author[add1]{Il-Chul Moon}
\author[add5]{Weonyoung Joo\corref{cor}}
\ead{weonyoungjoo@ewha.ac.kr}

\address[add1]{Department of Industrial and Systems Engineering, Korea Advanced Institute of Science and Technology, Daejeon, Republic of Korea}
\address[add2]{Department of Artificial Intelligence, University of Seoul, Seoul, Republic of Korea}
\address[add3]{ioCrops, Seoul, Republic of Korea}
\address[add4]{Department of Mathematical Sciences, Korea Advanced Institute of Science and Technology, Daejeon, Republic of Korea}
\address[add5]{Department of Statistics, EWHA Womans University, Seoul, Republic of Korea}
\cortext[cor]{Corresponding author}

\begin{abstract}
Bayesian inference without the likelihood evaluation, or \textit{likelihood-free inference}, has been a key research topic in \textit{simulation} studies for gaining quantitatively validated simulation models on real-world datasets. As the likelihood evaluation is inaccessible, previous papers train the amortized neural network to estimate the ground-truth posterior for the simulation of interest. Training the network and accumulating the dataset alternatively in a sequential manner could save the total simulation budget by orders of magnitude. In the data accumulation phase, the new simulation inputs are chosen within a portion of the total simulation budget to accumulate upon the collected dataset so far. This newly accumulated data degenerates because the set of simulation inputs is hardly mixed, and this degenerated data collection process ruins the posterior inference. This paper introduces a new sampling approach, called Neural Proposal (NP), of the simulation input that resolves the biased data collection as it guarantees the i.i.d. sampling. The experiments show the improved performance of our sampler, especially for the simulations with multi-modal posteriors. 
\end{abstract}


\end{frontmatter}


\section{Introduction}\label{sec:Introduction}

In case of a rare event or a single event phenomenon, data collection is of the most important issue. Many disciplines of science, engineering, and economics rely on simulations as a data generation tool. A simulation imitates the real-world with a number of input parameters $\bm{\theta}$, which need to be adjusted to fit the simulation to the real-world. However, as a simulation is fundamentally a descriptive process, the likelihood function $p_{sim}(\mathbf{x}|\bm{\theta})$ is intractable in general. Given the implicit nature, the goal of \textit{likelihood-free inference} is estimating the posterior distribution $p_{sim}(\bm{\theta}\vert\mathbf{x}_{o})$ to validate our simulation to the real-world data, where the posterior provides the probability of simulation input $\bm{\theta}$ given a single data $\mathbf{x}=\mathbf{x}_{o}$.

Evidenced in many areas, such as cosmology \citep{shaw2007efficient}, biomechanics \citep{franck2017multimodal}, and geosciences \citep{lu2017bayesian}, \textit{likelihood-free inference} becomes particularly challenging when the posterior distribution $p_{sim}(\bm{\theta}|\mathbf{x})$ is multi-modal. This multi-modal assumption is satisfied if either of the following two conditions meets in practice. First, if the simulation is periodic for a specific parameter (e.g. the amplitude in the pendulum problem), then the posterior could be multi-modal unless the parameter search space is carefully selected. Second, if there are multiple candidates of the underlying mechanics for the observed summary statistics, the simulation could have numerous simulation inputs generating each of the candidate mechanics. The aforementioned multi-modal conditions often arise in the simulation studies \citep{townsend2021validation}, and we focus on this multi-modal posterior inference \citep{radev2020bayesflow}.

Recent trend of \textit{likelihood-free inference} is modeling either likelihood \citep{papamakarios2019sequential}, posterior \citep{greenberg2019automatic}, or the classifier \cite{hermans2019likelihood} with a neural network. Those models alternatively train their network and accumulate additional simulation input-output pairs to the training dataset, based on the trained network. This sequential approach is becoming the mainstream of \textit{likelihood-free inference} because it could save the simulation budget by orders of magnitude, where running the simulation is the most expensive building block of the inference task in general.

\begin{figure}[t]
	\centering
	\subfigure[Inferenced by SNL]{\includegraphics[width=.495\linewidth]{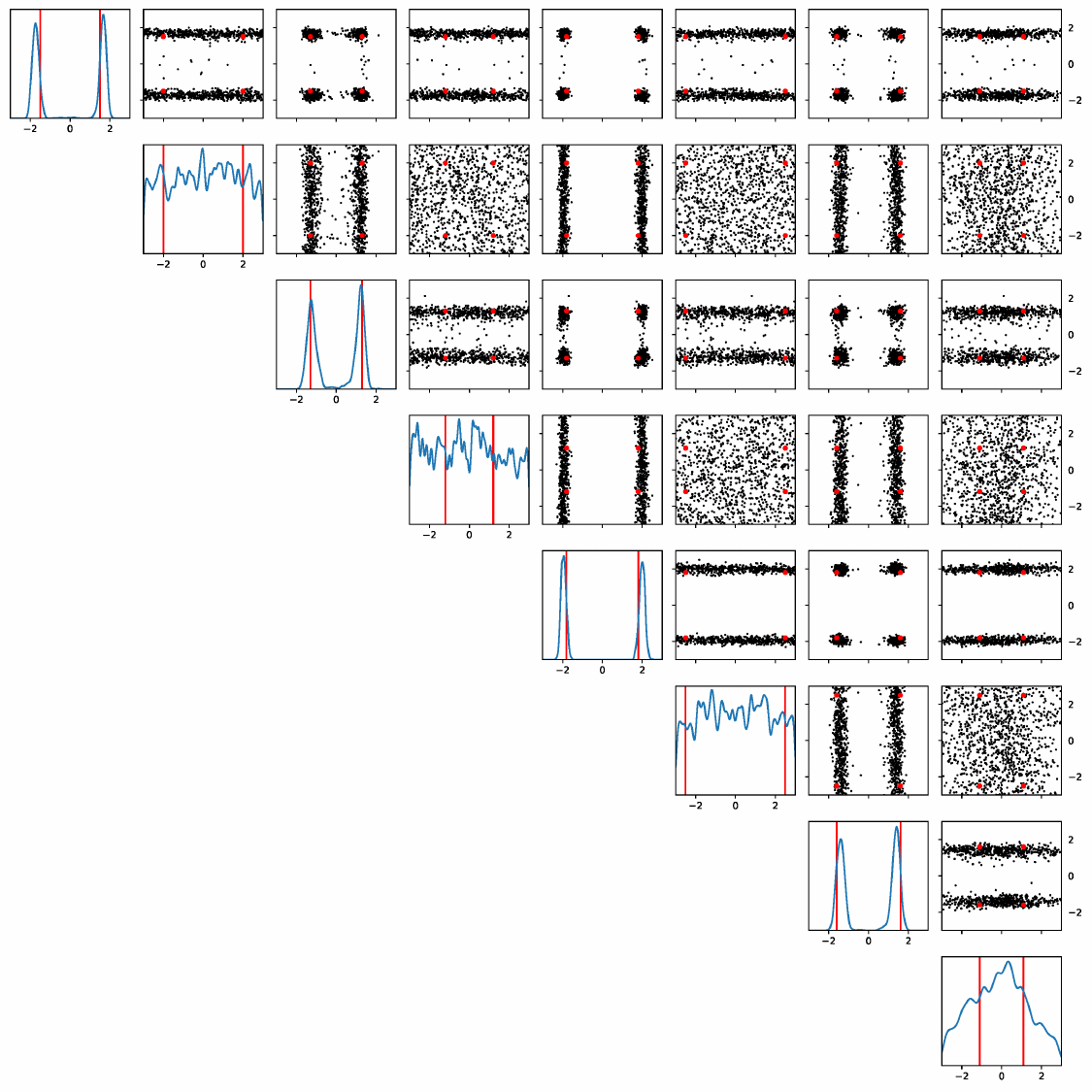}}
	\centering
	\subfigure[Inferenced by SNL with NP]{\includegraphics[width=.495\linewidth]{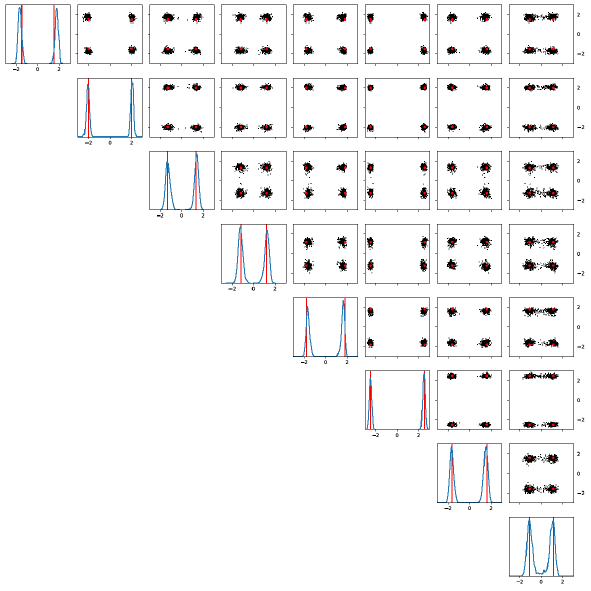}}
	\vskip -0.05in
	\caption{Comparison of samples from the approximate posteriors after $R$-rounds of inference by (a) Sequantial Neural Likelihood (SNL) \cite{papamakarios2019sequential} and (b) SNL with NP, on SLCP-256. We illustrate the samples from the approximate posterior via $1,000$-chained MCMC for both models. In the diagonal boxes, the marginal 1d empirical (blue) and true (red) distribution are plotted. In the off-diagonal boxes, the marginal 2d samples (blue) are visualized with the ground-truth values with red points. See Section \ref{sec:Experiments} for the details of SLCP-256.}
	\label{fig:Example}
	\vskip -0.05in
\end{figure}

Figure \ref{fig:Example}-(a), however, illustrates that the approximate posterior is unsuccessful with the current practice, if the posterior of interest is multi-modal. Figure \ref{fig:Example}-(a) happens not because of the training model but because the additional simulation inputs for model training degenerate. Therefore, we introduce a new sampling algorithm, Neural Proposal (NP), that is asymptotically nondegenerate. Figure \ref{fig:Example}-(b) visualizes the inference quality with the newly introduced NP jointly combined with the previous inference algorithm, SNL. Our contributions are summarized in the followings. (1) We propose a new sampler, Neural Proposal, for the balanced sampling that saves the computational budget when the posterior is multi-modal. (2) We theoretically show that the Neural Proposal is highly accurate to the original proposal distribution if the neural network is flexible enough.

\begin{figure*}[t]
	\centering
	\subfigure[Conventional SLFI with MCMC/AL.]{\includegraphics[width=.41\linewidth]{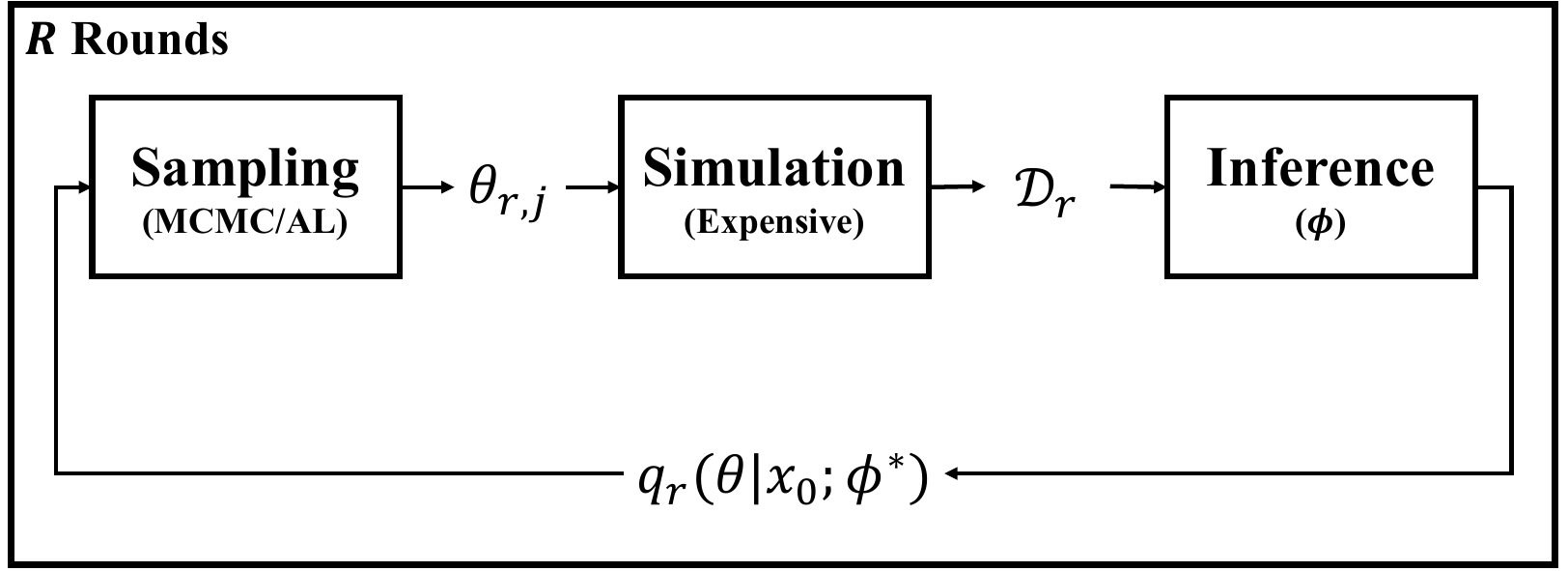}}
	\hspace{1.5em}
	\subfigure[Proposed SLFI with NP.]{\includegraphics[width=.41\linewidth]{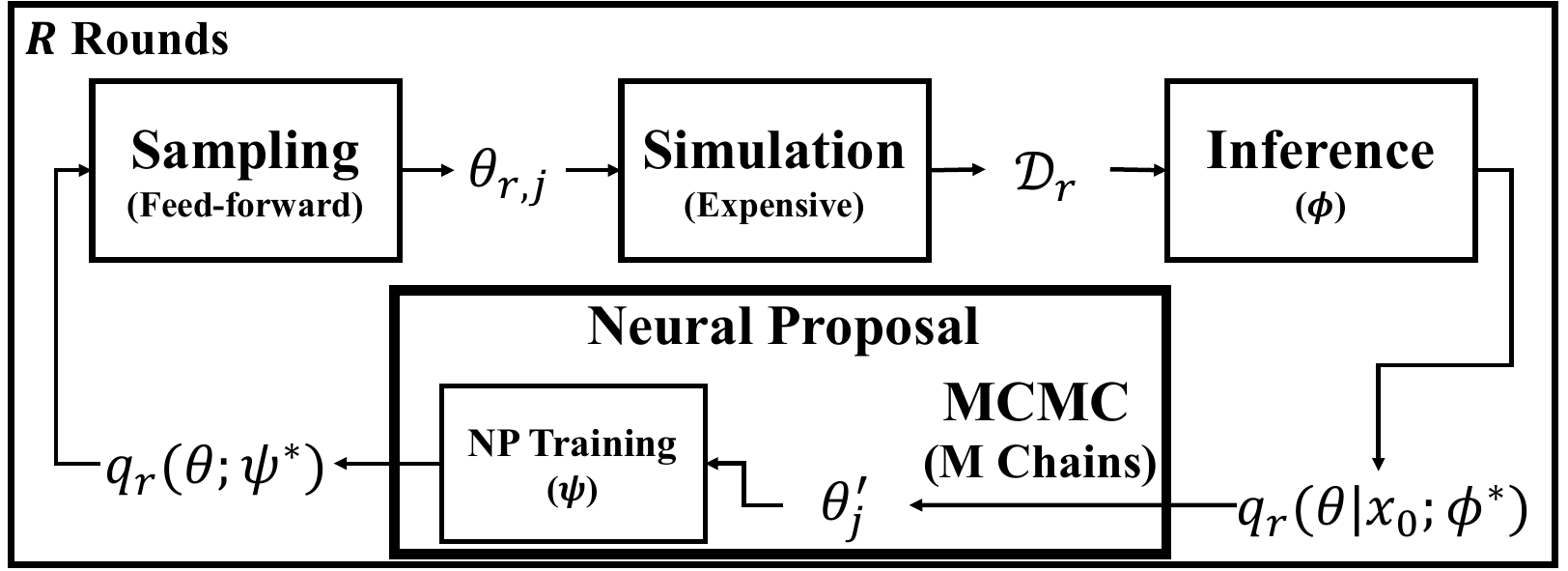}}
	\vskip -0.1in
	\caption{Component-wise algorithmic structures of (a) SLFI up to now and (b) SLFI with NP.}
	\label{fig:SLFI}
	\vskip -0.05in
\end{figure*}

\section{Preliminary}\label{sec:Preliminary}

\subsection{Problem Formulation}

A simulation is likelihood-free, meaning that only a sampling from $p_{sim}(\mathbf{x}\vert\bm{\theta})$ is accessible. The objective of \textit{likelihood-free inference} is estimating the true posterior distribution, $p_{sim}(\bm{\theta}|\mathbf{x}=\mathbf{x}_{o})$, with $N$ simulation budget for each round, given 1) the prior distribution $p(\bm{\theta})$ on the simulation input space, and 2) a single-shot real-world observation $\mathbf{x}_{o}$.

\subsection{Sequential Likelihood-Free Inference}\label{sec:SLFI}

Sequential Likelihood-Free Inference (SLFI) \citep{papamakarios2016fast} iteratively approximates the target posterior $p_{sim}(\bm{\theta}\vert\mathbf{x})$ with a modeled posterior $q_{r}(\bm{\theta}\vert\mathbf{x};\phi)$, parametrized by $\phi$, as in Figure \ref{fig:SLFI}-(a). These recurrent rounds fasten the approximate posterior $q_{r}$ to the true posterior $p_{sim}$ as round proceeds. In the \textit{sampling} step, few research \citep{lueckmann2019likelihood, durkan2018sequential} proposed to draw adaptive new inputs with active learning. Otherwise, the new simulation inputs are selected from the proposal distribution with MCMC. The proposal distribution is the latest approximate posterior $q_{r}(\bm{\theta}\vert\mathbf{x}=\mathbf{x}_{o};\phi^{*})$ at the observation $\mathbf{x}_{o}$, and this is based on the belief that the latest approximate posterior is the closest to the true posterior $p_{sim}(\bm{\theta}\vert\mathbf{x}_{o})$. As the approximate posterior converges to the true posterior, the proposal sampling is more likely to sample next $\bm{\theta}$s near the peaks of the true posterior which could enhance the inference quality with details near the peaks. After constructing a new dataset of $\mathcal{D}_{r}\leftarrow\mathcal{D}_{r-1}\cup\{(\bm{\theta}_{r,j},\mathbf{x}_{r,j})\}_{j=1}^{N}$ with $\mathbf{x}_{r,j}\sim p_{sim}(\cdot\vert\bm{\theta}_{r,j})$ by $N$ number of simulation runs, we train the neural network in the \textit{inference} stage with $\mathcal{D}_{r}$. A non-exhaustive list of previous works includes Sequential Neural Likelihood (SNL) \cite{papamakarios2019sequential}, Amortized Approximate Likelihood Ratio (AALR) \citep{hermans2019likelihood}, and Automatic Posterior Transformation (APT) \citep{greenberg2019automatic}. Table \ref{tab:previous} summarizes the estimation target and the sampling distribution of these previous algorithms.

\section{Motivation of a New Sampler}

\subsection{Empirical Motivation}\label{sec:empirical_motivation}

\begin{figure}[t]
	\centering
	\includegraphics[width=.7\linewidth]{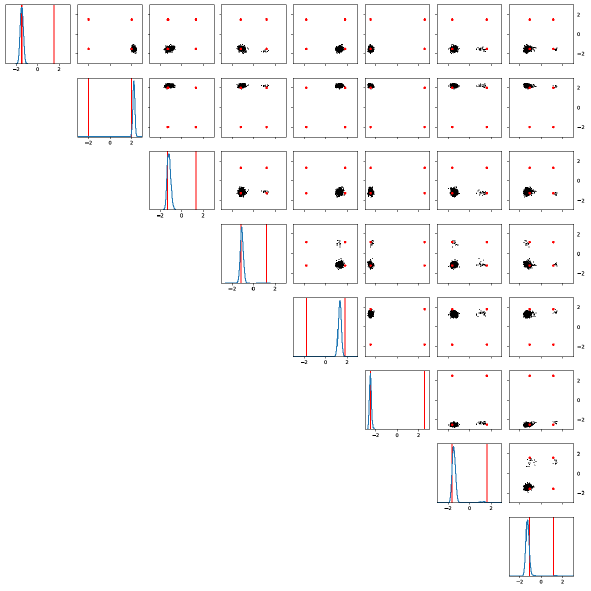}
	\vskip -0.1in
	\caption{Samples from MCMC with a single chain.}
	\label{fig:single_chain}
	\vskip -0.05in
\end{figure}

In practice, SNL and AALR draw additional $N$ simulation inputs at each round via MCMC as their sampling distribution is unnormalized, see Table \ref{tab:previous}. However, if the proposal distribution $q_{r}(\bm{\theta}\vert\mathbf{x}_{o};\phi^{*})$ attains the vast plateau of zero probability between modes, the jump between modes seldomly happens with MCMC samplers. Figure \ref{fig:single_chain} visualizes the slow mixing of MCMC, and this mode collapse issue in the sampling process leads $\mathcal{D}_{r}$ to be degenerate. This degeneracy is the main source of the waste of the simulation budget. The sampling issue could be partially mitigated if we use the multi-start MCMC \citep{chowdhury2018parallel}. The multi-start MCMC casts multiple chains in an independent manner, so each of $M$ uncorrelated chains draws $N/M$ simulation inputs. Better samples could be selected from these uncorrelated chains because multi-start forces explore the sample space \citep{altekar2004parallel}.

We experiment on the multi-start MCMC with a hypothetical simulation model, called SLCP-256, which has 256 symmetric modalities in its posterior. We defer the detailed description of SLCP-256 to Section \ref{sec:Experiments}. We know the exact (unnormalized) posterior distribution of SLCP-256, so we apply Metropolis-Hastings algorithm \cite{chib1995understanding} to draw $N=1,000$ samples from the exact posterior. Table \ref{tab:parallel_vs_surrogate} compares the multi-start MCMC with various $M$. Table \ref{tab:parallel_vs_surrogate} empirically supports that the multi-start MCMC significantly resolves the mixing problem as $M$ increases. However, as $M$ cannot exceed $N$, this restriction on $M$ fundamentally limits the sampling accuracy. Meanwhile, the newly introduced Neural Proposal (NP) releases this restriction of $M\le N$, and Table \ref{tab:parallel_vs_surrogate} shows that NP outperforms MCMC.

\begin{table}[t]
	\centering
	\caption{Estimation target and sampling distribution of previous works.}
	\label{tab:previous}
	\scriptsize
	\resizebox{.8\linewidth}{!}{
		\begin{tabular}{lcc}
			\toprule
			& \multirow{2}{*}{\shortstack{Estimation Target\\of Neural Network}} & \multirow{2}{*}{\shortstack{Sampling Distribution\\(Possibly Unnormalized)}} \\
			&&\\\midrule
			SNL \cite{papamakarios2019sequential} & $p_{sim}(\mathbf{x}\vert\bm{\theta})$ & $q_{r}(\mathbf{x}_{o}\vert\bm{\theta};\phi^{*})p(\bm{\theta})$ \\\midrule
			AALR \citep{hermans2019likelihood} & $p_{sim}(\mathbf{x}\vert\bm{\theta})/\tilde{p}_{r}(\mathbf{x})$ & $p(\bm{\theta})d_{r}(\bm{\theta},\mathbf{x}_{o};\phi^{*})/\big(1-d_{r}(\bm{\theta},\mathbf{x}_{o};\phi^{*})\big)$ \\\midrule
			APT \citep{greenberg2019automatic} & $p_{sim}(\bm{\theta}\vert\mathbf{x})$ & $q_{r}(\bm{\theta}\vert\mathbf{x}_{o};\phi^{*})$ \\
			\bottomrule
		\end{tabular}
	}
\end{table}

\begin{table}[t]
	\centering
	\caption{Comparison of 1) the Neural Proposal trained with $M=10^4$ samples from $M=10^4$ Markov chains and 2) MCMC with a various number of chains on SLCP-256. We measure the sample imbalance by the total variation distance between the uniform vector of $(1/256,...,1/256)$ and the empirical sample distribution $(v_{1},...,v_{256})$, where $v_{i}$ represents the proportion of samples fall in the $i$-th mode. We evaluate the effective sample size as $N\vert\Lambda\vert/\vert\Sigma\vert$, where $\Lambda$ and $\Sigma$ are the covariances of the proposal and the empirical distributions, respectively.}
	\label{tab:parallel_vs_surrogate}
	\scriptsize
	\resizebox{.9\linewidth}{!}{
		\begin{tabular}{cccccc}
			\toprule
			\multirow{2}{*}{\shortstack{Qualitative\\Metrics}} & \multicolumn{4}{c}{Proposal with MCMC} & \multirow{2}{*}{\shortstack{Neural\\Proposal}}\\\cmidrule(lr){2-5}
			&$M=1$&10&100&1,000&\\\midrule
			\multirow{2}{*}{\shortstack{Missed\\Mode ($\downarrow$)}} & 248 & 229 & 157 & 14 & \textbf{7}\\
			& ($\pm$2) & ($\pm$3) & ($\pm$4) & ($\pm$3) & ($\pm$2) \\\midrule
			\multirow{2}{*}{\shortstack{Sample\\Imbalance ($\downarrow$)}} & 1.93 & 1.79 & 1.30 & 0.55 & \textbf{0.44}\\
			& ($\pm$0.03) & ($\pm$0.02) & ($\pm$0.03) & ($\pm$0.03) & ($\pm$0.02) \\\midrule
			\multirow{2}{*}{\shortstack{Effective\\Sample Size ($\uparrow$)}} & 25 & 205 & 779 & 981 & \textbf{995} \\
			& ($\pm$22) & ($\pm$82) & ($\pm$52) & ($\pm$36) & ($\pm$23) \\
			\bottomrule
		\end{tabular}
	}
\end{table}

\subsection{Theoretical Motivation}\label{sec:theoretic_motivation}

In this section, we prove that the \textit{sampling error} is indeed strictly decreasing by $M$ which is consistent with the result in Table \ref{tab:parallel_vs_surrogate}. To define the \textit{sampling error}, suppose $P_{r}(A,\bm{\theta})$ is the transition probability of MCMC from $\bm{\theta}$ to a measurable set $A$ at the $r$-th round. Then, the probability distribution after $t$-Markov transition initialized at $\pi$ becomes $(\pi P_{r}^{t})(A):=\mathbb{E}_{\pi(\bm{\theta})}[P_{r}^{t}(A,\bm{\theta})]$, and this probability converges to the proposal distribution $\pi P_{r}^{t}\rightarrow q_{r}(\cdot\vert\mathbf{x}_{o};\phi^{*})$ as $t\rightarrow\infty$ for an ergodic Markov chain. This means that MCMC could randomly remain the search space on simulation input from one mode to another arbitrary mode if we transit the particle infinitely many times. However, this convergence theory is highly impractical in the practice for its extremely slow mixing rate, as depicted in Figure \ref{fig:single_chain}.

Therefore, instead of analyzing the  convergence rate of a single chain, we analyze the \textit{sampling error} by $M$. Having the initial points $\{\bm{\theta}_{j}\}_{j=1}^{M}$ sampled from $\pi$, the sampling distribution becomes $\frac{1}{M}\sum_{j=1}^{M}P_{r}^{t}(A,\bm{\theta}_{j})$, and we define the \textit{sampling error} of multi-chain MCMC as
\begin{align}\label{eq:sampling_error}
\text{e}_{MCMC}(M)=\mathbb{E}_{\{\bm{\theta}_{j}\}_{j=1}^{M}\sim\pi}\bigg[D\bigg(q_{r}(\cdot\vert\mathbf{x}_{o};\phi^{*})\Big\Vert\frac{1}{M}\sum_{j=1}^{M}P_{r}^{t}(\cdot,\bm{\theta}_{j})\bigg)\bigg],
\end{align}
where $D$ is an arbitrary divergence. Then, Theorem \ref{prop:monotone} clarifies the connection between the sampling error of MCMC with respect to $M$. According to Theorem \ref{prop:monotone}, $\text{e}_{MCMC}$ remains to be strictly positive if $M\le N$. This theoretic analysis with the empirical study in Section \ref{sec:empirical_motivation} highly motivates a new sampler that is free from the restriction of $M\le N$. 
\begin{theorem}\label{prop:monotone}
    Suppose a divergence $D$ is convex in its second argument\footnotemark[2]. Then, the \textit{sampling error} $\text{e}_{MCMC}(M)$ strictly decreases by $M$. Additionally, if $D$ induces a weak topology\footnotemark[3], $\text{e}_{MCMC}(M)$ converges to zero as $M\rightarrow\infty$ and $t\rightarrow\infty$.
\end{theorem}\footnotetext[2]{Any type of the Integral Probability Metric \citep{sriperumbudur2010hilbert} and the KL-divergence satisfy this convexity condition.}\footnotetext[3]{Large family of Integral Probability Metrics, such as the Wasserstein metric, the maximum mean divergence, the Prohorov metric, and the Dudley metric induce the weak topology \citep{sriperumbudur2010hilbert}.}
\begin{proof}
We prove a more general statement that
\begin{align*}
	\mathbb{E}_{\mathbb{P}}\Big[D\Big(Q\big\Vert P_{M}^{t}(\bm{\bm{\theta}}_{1:\infty})\Big)\Big]\le\mathbb{E}_{\mathbb{P}}\Big[D\Big(Q\big\Vert P_{M'}^{t}(\bm{\bm{\theta}}_{1:\infty})\Big)\Big],\tag{$*$}
	\end{align*}
	for any distribution $Q$ and $M\ge M'$, where $P_{M}^{t}(\bm{\theta}_{1:\infty})(A):=\frac{1}{M}\sum_{j=1}^{M}P_{r}^{t}(A,\bm{\theta}_{j})$ and $(\text{supp}(\pi)^{\infty},\mathcal{F},\mathbb{P})$ is the probability space from the Kolmogorov extention theorem, such that $\mathbb{E}_{\bm{\theta}_{j}\sim\pi}[f(\bm{\theta}_{j})]=\mathbb{E}_{\bm{\theta}_{1:\infty}\sim\mathbb{P}}[f(\bm{\theta}_{1:M})]$, for any measurable function $f$ defined on $\text{supp}(\pi)$. 
    Suppose Equation ($*$) holds for $M' = M-1$ and any $M\ge 1$. Then, the desired result holds for any $M\ge M'$ if we apply the above inequality multiple times. Now, 
	\begin{align*}
	&\mathbb{E}_{\mathbb{P}}\Big[D\Big(Q\big\Vert P_{M}^{t}(\bm{\bm{\theta}}_{1:\infty})\Big)\Big]=\mathbb{E}_{\mathbb{P}}\Big[D\Big(Q\big\Vert \frac{1}{M}\sum_{k=1}^{M}\frac{1}{M-1}\sum_{\substack{j=1, j\ne k}}^{M}P_{r}^{t}(\cdot,\bm{\bm{\theta}}_{j})\Big)\Big]\\
	&\quad\quad\quad\quad\quad\quad\quad\le\frac{1}{M}\sum_{k=1}^{M}\mathbb{E}_{\mathbb{P}}\Big[D\Big(Q\big\Vert\frac{1}{M-1}\sum_{\substack{j=1, j\ne k}}^{M}P_{r}^{t}(\cdot,\bm{\bm{\theta}}_{j})\Big)\Big]\\
	&\quad\quad\quad\quad\quad\quad\quad=\mathbb{E}_{\mathbb{P}}\Big[D\Big(Q\big\Vert P_{M-1}^{t}(\bm{\bm{\theta}}_{1:\infty})\Big)\Big],
	\end{align*}
	where the inequality holds from the Jensen's inequality and the equality in the last line holds since the random variable of $\sum_{\substack{j=1, j\ne k}}^{M}P_{r}^{t}(\cdot,\bm{\bm{\theta}}_{j})$ equals to that of $\sum_{j=1}^{M-1}P_{r}^{t}(\cdot,\bm{\bm{\theta}}_{j})$ in distribution for any $k=1,...,M$.
	
	The equality condition of the above inequality is $P_{r}^{t}(\cdot,\bm{\bm{\theta}}_{j})=P_{r}^{t}(\cdot,\bm{\bm{\theta}}_{j'})$ for all $j,j'=1,...,M$. However, if the initial state does not match, then $P_{r}^{t}$ would not be equal each other unless $P_{r}^{t}$ is degenerate. Since the neural network is continuous, the transition probability of any MCMC would be continuous, and $P_{r}^{t}$ would not be degenerate.
\end{proof}

\section{Neural Proposal}\label{sec:Methodology}

Neural Proposal releases the restriction of $M\le N$ by \textit{not} using MCMC samples directly as the simulation inputs. Instead, Neural Proposal is a neural sampler that replaces the unnormalized proposal distribution, and we select the next batch of simulation inputs from this neural sampler. We construct the training dataset for this neural sampler by the samples from the proposal distribution. Hence, we could increase the number of Markov chains as much as we desire. After the training, the simulation inputs for the next round are sampled via the feed-forward computation, and this feed-forward sampling naturally implies the i.i.d-ness of the Neural Proposal.

\begin{figure}[t]
    \centering
    \includegraphics[width=0.6\linewidth]{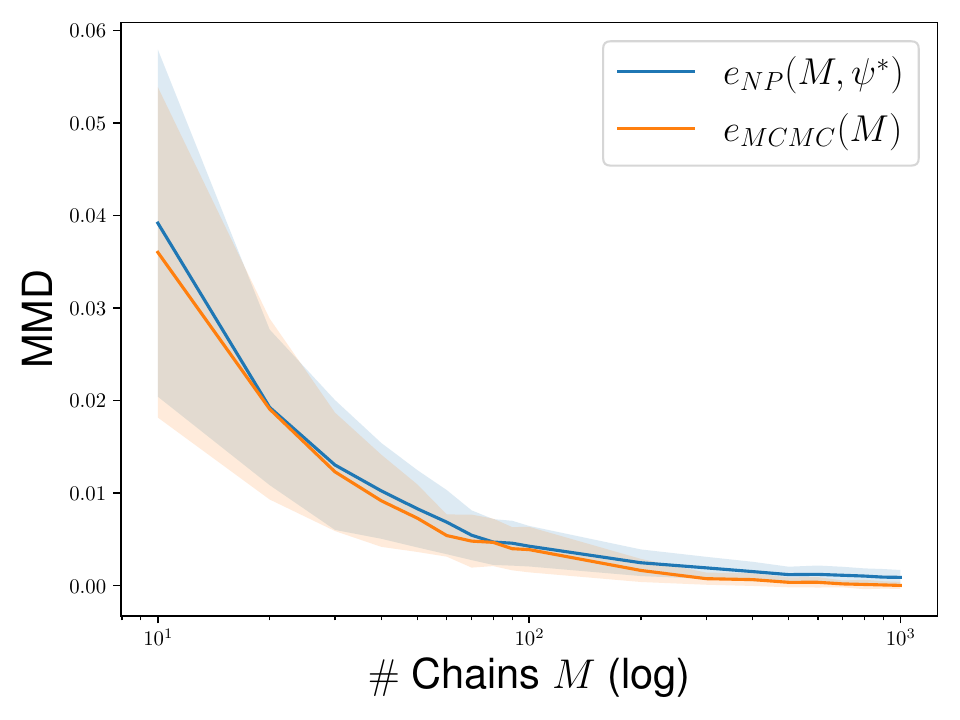}
    \vskip -0.05in
    \caption{Empirical study on the errors $e_{NP}(M,\psi^{*})$ and $e_{MCMC}(M)$, measured with Maximum Mean Discrepancy (MMD) \cite{sriperumbudur2010hilbert}.}
    \label{fig:convergence}
    \vskip -0.05in
\end{figure}

Concretely, we model the Neural Proposal $q_{r}(\bm{\theta};\psi)$ with a normalizing flow, parametrized by $\psi$, with the loss function
\begin{align*}
\mathcal{L}_{NP}(\mathbf{\psi})=-\mathbb{E}_{\bm{\theta}'\sim (\pi P_{r}^{t})}[\log{q_{r}(\bm{\theta}';\psi)}],
\end{align*}
which is equivalent to
\begin{align}\label{eq:neural_posterior}
\begin{split}
\mathcal{L}_{NP}(\mathbf{\psi})&=-\frac{1}{M}\sum_{j=1}^{M}\mathbb{E}_{\bm{\theta}_{j}\sim\pi}\mathbb{E}_{\bm{\theta}_{j}'\sim P_{r}^{t}(\cdot, \bm{\theta}_{j})}[\log{q_{r}(\bm{\theta}_{j}';\psi)}]\\
&=\mathbb{E}_{\{\bm{\theta}_{j}\}_{j=1}^{M}\sim\pi}\bigg[D_{KL}\bigg( \frac{1}{M}\sum_{j=1}^{M}P_{r}^{t}(\cdot,\bm{\theta}_{j})\bigg\Vert q_{r}(\cdot;\psi) \bigg)\bigg],
\end{split}
\end{align}
up to a constant, if we use $M$ Markov chains to construct the training dataset for Neural Proposal. Combining Eq. \ref{eq:neural_posterior} with Eq. \ref{eq:sampling_error}, the \textit{sampling error} of Neural Proposal $e_{NP}(M,\psi)$ satisfies
\begin{align*}
e_{NP}(M,\psi):&=D\big(q_{r}(\cdot\vert\mathbf{x}_{o};\phi^{*})\big\Vert q_{r}(\cdot;\psi)\big)\\
&\le \mathbb{E}_{\{\bm{\theta}_{j}\}_{j=1}^{M}\sim\pi}\bigg[D\bigg(q_{r}(\cdot\vert\mathbf{x}_{o};\phi^{*})\Big\Vert\frac{1}{M}\sum_{j=1}^{M}P_{r}^{t}(\cdot,\bm{\theta}_{j})\bigg)\\
&\quad+D\bigg( \frac{1}{M}\sum_{j=1}^{M}P_{r}^{t}(\cdot,\bm{\theta}_{j})\bigg\Vert q_{r}(\cdot;\psi) \bigg)\bigg]\\
&\le\text{e}_{MCMC}(M)+\mathcal{L}_{NP}(\psi),
\end{align*}
if $D$ is a weaker metric than the KL divergence. Therefore, the optimal Neural Proposal has the \textit{sampling error} of
\begin{align}\label{eq:connection}
e_{NP}(M,\psi^{*})=D\big(q_{r}(\cdot\vert\mathbf{x}_{o};\phi^{*})\big\Vert q_{r}(\cdot;\psi^{*})\big)\le \text{e}_{MCMC}(M)+\epsilon,
\end{align}
where $\epsilon$ is the estimation error, i.e., $\mathcal{L}_{NP}(\psi^{*})\le \epsilon$. Indeed, Figure \ref{fig:convergence} illustrates that the inequality \ref{eq:connection} between $e_{NP}(M,\psi^{*})$ and $e_{MCMC}(M)$ is tight enough if we use a flexible neural network for NP, throughout the choice of $M$.

\begin{table*}[t]
	\renewcommand{\arraystretch}{1.4}
	\caption{Quantitative performances of synthetic simulation models with tractable likelihoods. The lower the better for NLTP/C2ST/log MMD. The higher the better for IS. The boldface numbers perform the best out of the corresponding column. FF denotes the feed-forward sampler.}
	\label{tab:performance}
	\vskip -0.05in
	\scriptsize
	\resizebox{.95\linewidth}{!}{
		\begin{threeparttable}{
				\begin{tabular}{cccrrrrrrrr}
					\toprule
					\multirow{2}{*}{Algorithm} & \multirow{2}{*}{Sampling} & \multirow{2}{*}{\shortstack{Sampler\\($\#$ Chains)}} & \multicolumn{4}{c}{SLCP-16} & \multicolumn{4}{c}{SLCP-256}\\\cmidrule(lr){4-11}
					& & & \multicolumn{1}{c}{NLTP ($\downarrow$)} & \multicolumn{1}{c}{C2ST ($\downarrow$)} & $\log{\text{MMD}}$ ($\downarrow$) & \multicolumn{1}{c}{IS ($\uparrow$)} & \multicolumn{1}{c}{NLTP\tnote{*}} & \multicolumn{1}{c}{C2ST} & \multicolumn{1}{c}{$\log{\text{MMD}}$} & \multicolumn{1}{c}{IS}\\\midrule
					SMC-ABC & Empirical & --- & 139.02$\pm$4.21 & 0.79$\pm$0.03 & -2.56$\pm$1.68 & 8.87$\pm$0.39 & 38.80$\pm$0.41 & 0.91$\pm$0.07 & -2.43$\pm$0.18 & 183.72$\pm$8.15\\\midrule
					APT & Direct & FF & 5.97$\pm$12.05 & 0.77$\pm$0.03 & -1.83$\pm$0.85 & 13.27$\pm$2.04 & 14.78$\pm$1.09 & 0.79$\pm$0.06 & -6.02$\pm$0.64 & 221.25$\pm$10.06 \\\midrule
					\multirow{10}{*}{SNL} & \multirow{7}{*}{MCMC} & Slice (1) &10.24$\pm$7.09 & 0.81$\pm$0.04 & -1.09$\pm$0.03 & 13.88$\pm$0.95 & 24.70$\pm$5.50 & 0.85$\pm$0.02 & -1.52$\pm$0.32 & 187.34$\pm$17.56\\
					&  & Slice (10)\tnote{**} & 8.62$\pm$4.86 & 0.77$\pm$0.03 & -1.76$\pm$0.67 & 13.64$\pm$1.18 & 14.43$\pm$3.06 & 0.84$\pm$0.02 & -1.36$\pm$0.23 & 222.60$\pm$9.28\\\cline{3-11}
					&  & MH (1) & -11.04$\pm$25.34 & 0.78$\pm$0.02 & -0.99$\pm$0.03 & 15.76$\pm$0.33 & 23.94$\pm$8.63 & 0.86$\pm$0.02 & -1.01$\pm$0.05 & 209.47$\pm$15.58\\
					&  & MH (100) & -18.01$\pm$13.47 & 0.75$\pm$0.03 & -5.71$\pm$0.69 & 15.72$\pm$0.47 & 17.38$\pm$8.51 & 0.83$\pm$0.04 & -5.62$\pm$0.47 & 217.44$\pm$12.81\\
					& & MH (1,000) & -30.28$\pm$15.87 & 0.74$\pm$0.03 & -6.12$\pm$2.65 & 15.70$\pm$0.42 & 16.90$\pm$8.95 & 0.79$\pm$0.07 & -6.14$\pm$1.28 & 212.22$\pm$21.44 \\\cline{3-11}
					&  & NUTS (1) & -23.43$\pm$5.14 & 0.80$\pm$0.05 & -0.98$\pm$0.05 & 15.75$\pm$0.38 & 22.19$\pm$3.51 & 0.87$\pm$0.04 & -0.98$\pm$0.06 & 209.89$\pm$9.17 \\
					& & NUTS (100) & -24.77$\pm$15.21 & 0.75$\pm$0.02 & -4.25$\pm$0.24 & 15.84$\pm$0.01 & 23.16$\pm$6.62 & 0.81$\pm$0.03 & -3.88$\pm$1.38 & 205.98$\pm$15.96\\\cmidrule(lr){2-11}
					& \multirow{3}{*}{\shortstack{Active\\Learning}} & MaxVar & -13.43$\pm$8.06 & 0.75$\pm$0.06 & -3.39$\pm$0.14 & 15.79$\pm$0.02 & 19.15$\pm$2.64 & 0.82$\pm$0.02 & -5.69$\pm$0.24 & 206.84$\pm$8.17 \\
					&  & MaxEnt & -21.19$\pm$26.16 & 0.76$\pm$0.01 & -3.52$\pm$0.16 & 15.72$\pm$0.38 & 22.24$\pm$6.61 & 0.80$\pm$0.01 & -5.72$\pm$0.22 & 202.74$\pm$9.41 \\
					&  & MaxBALD & -20.59$\pm$18.43 & 0.73$\pm$0.02 & -3.49$\pm$0.21 & 15.65$\pm$0.32 & 22.29$\pm$4.53 & 0.83$\pm$0.03 & -5.47$\pm$0.18 & 202.34$\pm$6.21\\\midrule
					\multirow{5}{*}{AALR} & \multirow{5}{*}{MCMC} & Slice (1) & 65.09$\pm$9.44 & 0.82$\pm$0.03 & -1.48$\pm$0.74 & 10.97$\pm$0.61 & 21.25$\pm$4.07 & 0.85$\pm$0.03 & -3.37$\pm$0.03 & 209.50$\pm$10.61\\
					&  & Slice (10) & 49.17$\pm$12.52 & 0.78$\pm$0.03 & -2.79$\pm$0.43 & 12.02$\pm$0.76 & 16.56$\pm$1.28 & 0.85$\pm$0.01 & -3.47$\pm$0.07 & 215.75$\pm$12.90\\\cline{3-11}
					&  & MH (1) & 50.01$\pm$20.49 & 0.82$\pm$0.03 & -1.11$\pm$0.15 & 12.55$\pm$1.18 & 22.82$\pm$4.49 & 0.86$\pm$0.02 & -1.01$\pm$0.05 & 197.03$\pm$23.96\\
					&  & MH (100) & 42.42$\pm$28.07 & 0.80$\pm$0.01 & -3.83$\pm$0.67 & 14.03$\pm$0.91 & 19.97$\pm$2.18 & 0.84$\pm$0.01 & -5.62$\pm$0.47 & 215.78$\pm$9.69\\
					& & MH (1,000) & 39.98$\pm$30.37 & 0.77$\pm$0.02 & -3.80$\pm$0.43 & 13.14$\pm$1.24 & 26.24$\pm$3.07 & 0.82$\pm$0.04 & -5.13$\pm$1.26 & 218.29$\pm$4.20 \\
					\midrule
					SNL & \multirow{2}{*}{\shortstack{Neural\\Proposal}} & \multirow{2}{*}{FF (10,000)} & \textbf{-32.11}$\pm$19.70 & \textbf{0.72}$\pm$0.04 & \textbf{-7.02}$\pm$0.95 & \textbf{15.85}$\pm$0.29 & \textbf{12.30}$\pm$2.58 & \textbf{0.77}$\pm$0.03 & \textbf{-6.87}$\pm$0.64 & \textbf{223.77}$\pm$12.22\\
					AALR &  &  & 31.56$\pm$19.43 & 0.78$\pm$0.01 & -3.68$\pm$0.53 & 12.95$\pm$1.69 & 18.28$\pm$2.54 & 0.83$\pm$0.01 & -5.78$\pm$1.18 & 220.00$\pm$8.38\\
					\bottomrule
				\end{tabular}
				\begin{tablenotes}
					\item[*] Scaled by $10^{-2}$
					\item[**] It is intractable to increase the number of chains for the slice and the NUTS samplers up to $M=1,000$ due to the computational bottleneck.
			\end{tablenotes}}
		\end{threeparttable}
	}
	\vskip -0.1in
\end{table*}

\begin{table}[t]
	\renewcommand{\arraystretch}{1.2}
	\centering
	\caption{NLTP of real-world simulation models.}
	\label{tab:performance_real}
	\vskip -0.05in
	\scriptsize
	\resizebox{.9\linewidth}{!}{
		\begin{tabular}{cccrr}
			\toprule
			\multirow{2}{*}{Algorithm} & \multirow{2}{*}{Sampling} & \multirow{2}{*}{\shortstack{Sampler\\($\#$ Chains)}} & \multicolumn{1}{c}{M/G/1} & \multicolumn{1}{c}{CLV}\\\cline{4-5}
			& & & \multicolumn{1}{c}{NLTP} & \multicolumn{1}{c}{NLTP} \\\midrule
			SMC-ABC & Empirical & --- & 3.89$\pm$1.10 & 16.77$\pm$4.60 \\\midrule
			APT & Direct & FF & -0.48$\pm$0.95 & 10.05$\pm$3.60\\\midrule
			\multirow{4}{*}{SNL} & \multirow{4}{*}{MCMC} & Slice (1) & -0.99$\pm$0.29 & 12.62$\pm$1.98 \\
			&  & Slice (10) & -0.85$\pm$0.41 & 12.34$\pm$1.09 \\\cline{3-5}
			&  & MH (1) & -0.92$\pm$0.58 & 29.36$\pm$9.27 \\
			&  & MH (100) & -1.01$\pm$0.32 & 21.99$\pm$8.30 \\\midrule
			\multirow{4}{*}{AALR} & \multirow{4}{*}{MCMC} & Slice (1) & -0.62$\pm$0.26 & 14.95$\pm$5.25 \\
			&  & Slice (10) & -0.38$\pm$0.30 & 16.80$\pm$6.25 \\\cline{3-5}
			&  & MH (1) & -0.64$\pm$0.36 & 26.88$\pm$6.78 \\
			&  & MH (100) & -0.57$\pm$0.68 & 17.03$\pm$5.72 \\\midrule
			SNL & \multirow{2}{*}{\shortstack{Neural\\Proposal}} & \multirow{2}{*}{FF (10,000)} & \textbf{-1.12}$\pm$0.50 & \textbf{9.69}$\pm$4.11 \\\cline{1-1}
			AALR &  &  & -0.60$\pm$0.43 & 11.83$\pm$1.79 \\
			\bottomrule
		\end{tabular}
	}
	\vskip -0.05in
\end{table}

Figure \ref{fig:SLFI}-(b) shows the component-wise sequential likelihood-free inference with NP as described below:
\begin{enumerate}[itemsep=0.5pt]
    \item \textbf{Sampling} Draw $N$ \textit{feed-forward} simulation inputs of $\{\bm{\theta}_{r,j}\}_{j=1}^{N}$ from the Neural Proposal $q_{r}(\cdot;\psi^{*})$.
    \item \textbf{Simulation} Simulate with $\bm{\theta}_{r,j}$ and add the simulated results to the dataset $\mathcal{D}_{r}\leftarrow\mathcal{D}_{r-1}\cup\{(\bm{\theta}_{r,j},\mathbf{x}_{r,j})\}_{j=1}^{N}$
    \item \textbf{Inference} Train the approximate posterior by $\phi^{*}=\argmin_{\phi}q_{r+1}(\bm{\theta}\vert\mathbf{x};\phi)$ with the new dataset $\mathcal{D}_{r}$.
    \item \textbf{Neural Proposal}
    \begin{enumerate}
        \item Draw $M$ samples from $q_{r}(\cdot\vert\mathbf{x}_{o},\phi^{*})$ via MCMC with $M$-chains.
        \item Train the Neural Proposal by $\psi^{*}=\argmin_{\psi}\mathcal{L}_{NP}(\psi)$ with $M$ MCMC samples.
    \end{enumerate}
\end{enumerate}


\section{Experiments}\label{sec:Experiments}

We compare our algorithm with SNL \citep{papamakarios2019sequential}, AALR \citep{hermans2019likelihood}, APT \citep{greenberg2019automatic}, and SMC-ABC \citep{sisson2007sequential}. We apply NP in place of SNL's and AALR's sampling algorithms for the next simulation inputs. We do not use NP in APT because APT already draws its samples of simulation inputs without MCMC by its algorithmic design. We use Neural Spline Flows \cite{durkan2019neural} for modeling both Neural Proposal and inference distributions. The implementation including the experiments can be found at \url{https://github.com/Kim-Dongjun/Neural_Proposal}.

\paragraph{Sampling Baselines} We compare the suggested NP against the previous MCMCs, including the slice sampler \citep{neal2003slice}, the Metropolis-Hastings (MH) sampler \citep{chib1995understanding}, the No-U-Turn-Sampler (NUTS) \citep{hoffman2014no}, and the active learning approaches \citep{lueckmann2019likelihood}. Among the active learning approaches, the maximum variance selects the next inputs by maximizing the variance of $q_{r}(\mathbf{x}_{o}\vert\bm{\theta};\phi_{k}^{*})$, where the variance is computed by the ensemble of multiple networks $\phi_{k}^{*}$, trained independently. Likewise, entropy-based active learning utilizes entropy to choose the next inputs. Similarly, we test the active learning with BALD \cite{houlsby2011bayesian} activation function.

\paragraph{Simulation Setting} Motivated from \citet{papamakarios2019sequential}, we propose the variants of Simple-Likelihood-and-Complex-Posterior (SLCP) simulation, and name those by SLCP-16 and SLCP-256. These generalized SLCP-family are designed to test the inference algorithms with more complex posteriors on high-dimensional simulation outputs. The governing equations of SLCP-16 is
\begin{align*}
&\quad\theta_{i}\sim U(-3,3) \quad \text{for} \quad i=1,...,5\\
&\quad \mathbf{m}_{\bm{\bm{\theta}}}=(\theta_{1}^{2},\theta_{2}^{2})\\
&\quad s_{1}=\theta_{3}^{2},\quad s_{2}=\theta_{4}^{2},\quad \rho=\tanh{(\theta_{5})}\\
&\quad \mathbf{S}_{\bm{\bm{\theta}}}=
\begin{pmatrix}
s_{1}^{2} & \rho s_{1}s_{2}\\
\rho s_{1}s_{2} & s_{2}^{2}
\end{pmatrix}\\
&\quad \mathbf{x}=\mathbf{x}_{1}\oplus\cdots\oplus\mathbf{x}_{25}\quad\text{with}\quad \mathbf{x}_{j}\sim \mathcal{N}(\mathbf{m_{\bm{\theta}}},\mathbf{S_{\bm{\theta}}}),
\end{align*}
where $\oplus$ notation denotes the vector concatenation. Overall, the simulation input is five-dimensional, and the output is 50-dimensional. We select the true input as $\bm{\theta}^{*}=(1.5,-2.0,-1.0,-0.9,0.6)$, but there are 15 alternative inputs that generate the identical simulation result (in distribution) with $(\pm1.5,\pm2.0,\pm1.0,\pm0.9,0.6)$. The prior is the uniform distribution on $[-3,3]^{5}$.

Analogous to SLCP-16, governing equations of SLCP-256 is
\begin{align*}
&\quad\theta_{i}\sim U(-3,3) \quad \text{for} \quad i=1,...,8\\
&\quad \mathbf{x}=\mathbf{x}_{1}\oplus\cdots\oplus \mathbf{x}_{5}\quad\text{where}\quad \mathbf{x}_{j}\sim \mathcal{N}(\bm{\theta}^{2},\mathbf{I}),
\end{align*}
with the simulation input/output to be 8/40-dimensions, respectively. The true input is $\bm{\theta}^{*}=(1.5,2.0,1.3,1.2,1.8,2.5,1.6,1.1)$ and there are 255 alternative optimal modes of those are symmetric to the origin. We search the optimal parameter from the uniform distribution on $[-3,3]^{8}$.

On top of these synthetic models, we test our Neural Proposal on M/G/1 and Competitive Lotka Volterra (CLV) \citep{vano2006chaos} simulations. The M/G/1 simulation is the most widely experimented simulation in the previous researches \citep{hermans2019likelihood, papamakarios2019sequential} with two dimensional inputs and five dimensional outputs. CLV is a generalization of the benchmark simulation on the Lotka-Volterra model \citep{papamakarios2019sequential, greenberg2019automatic}. Only the aggregated statistics is given as an observation in CLV, and this leads a bimodal posterior in CLV.

\begin{figure*}[t]
	\centering
	\includegraphics[width=\linewidth]{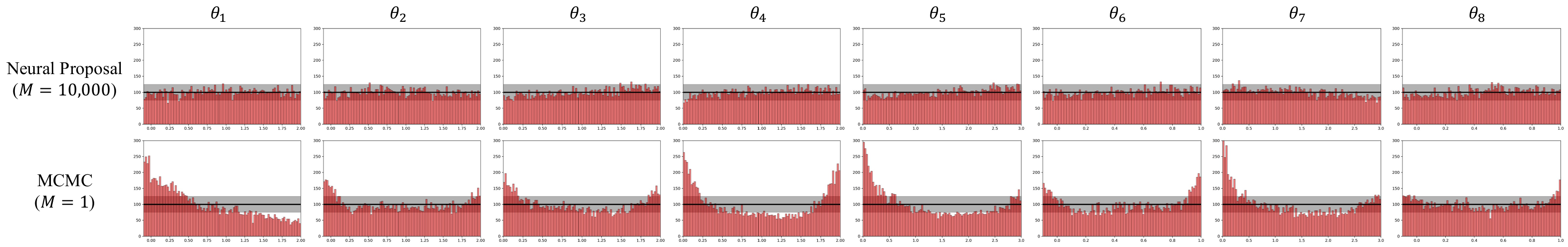}
	\vskip -0.05in
	\caption{Comparison of the inference quality of MCMC and NP on CLV. The histogram of NP is distributed more equally than that of MCMC, meaning the approximate posterior inferred with NP is more close to that of MCMC.}
	\label{fig:CLV}
	\vskip -0.1in
\end{figure*}

\begin{figure*}[t]
	\centering
	\subfigure[NLTP]{\includegraphics[width=.24\linewidth]{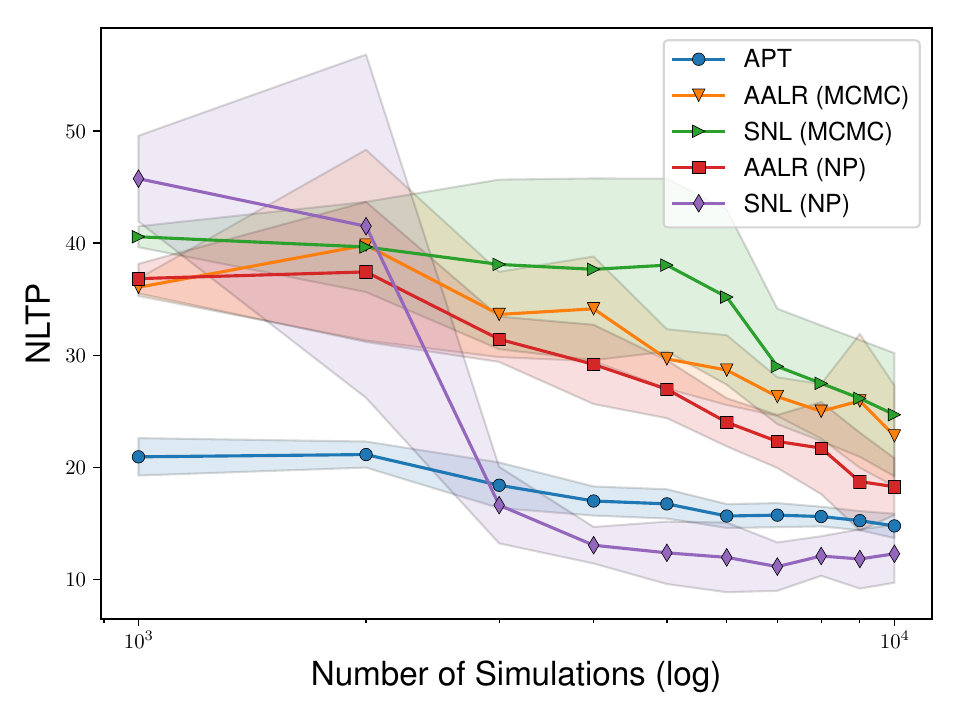}}
	\centering
	\subfigure[MMD]{\includegraphics[width=.24\linewidth]{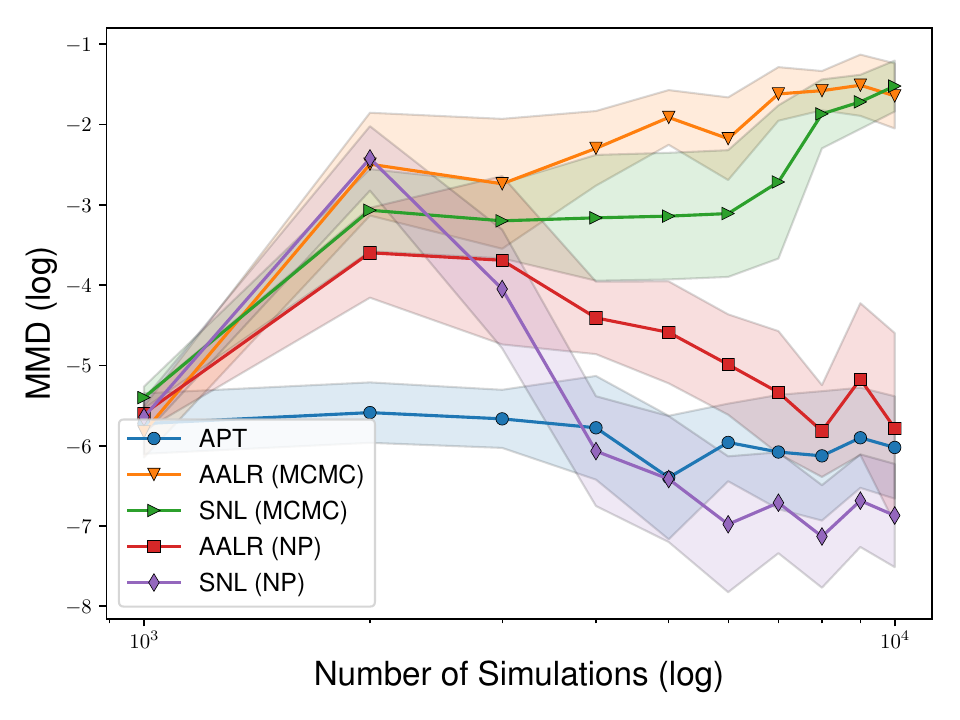}}
	\centering
	\subfigure[IS]{\includegraphics[width=.24\linewidth]{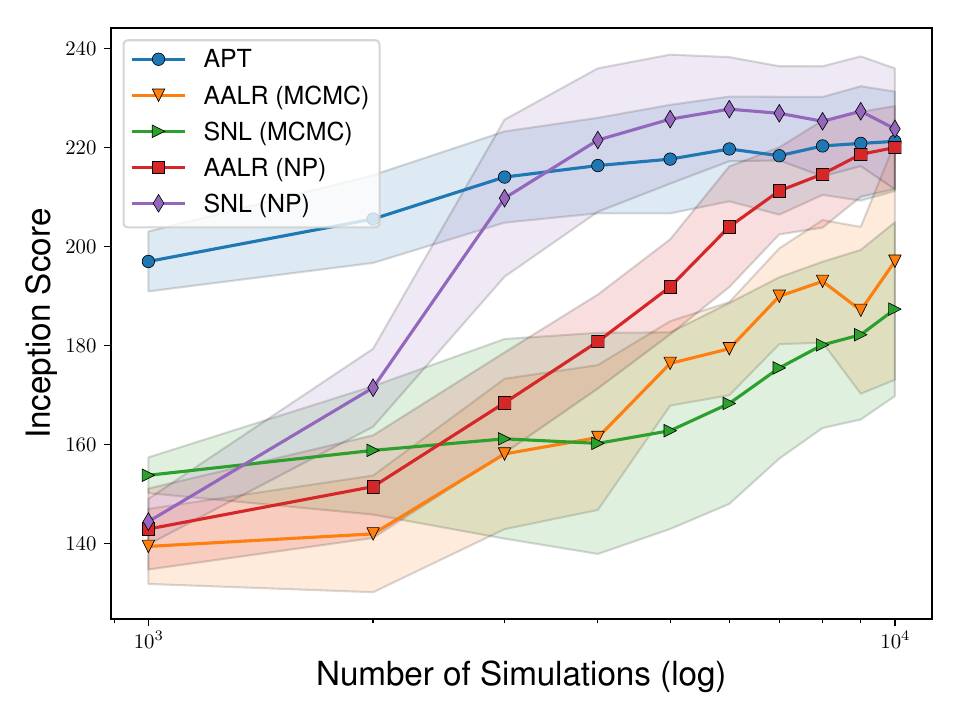}}
	\centering
	\subfigure[Missed Mode]{\includegraphics[width=.24\linewidth]{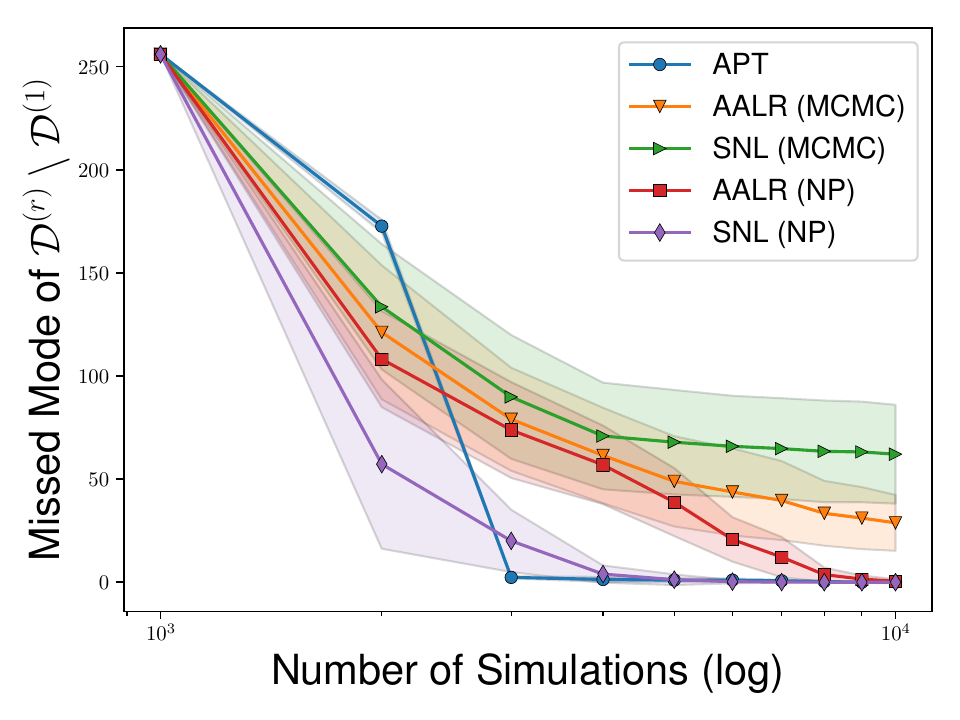}}
	\vskip -0.1in
	\caption{Comparison of baselines with the Neural Proposal trained with $M=10,000$ samples from $M=10,000$ Markov chains on SLCP-256 by round.}
	\label{fig:performance_by_round}
	\vskip -0.1in
\end{figure*}

\paragraph{Quantitative Measure} The default measure is the Negative Log-likelihood of True Parameters (NLTP), $-\sum_{i=1}^{n}\log{q_{R}(\bm{\theta}_{i}^{*}\vert\mathbf{x}_{o})}$ \citep{papamakarios2019sequential}, where $\{\bm{\theta}_{i}^{*}\}_{i=1}^{n}$ is the set of true and alternative parameters after the $R$ rounds of inference. However, NLTP ignores how the posterior is distributed other than $\bm{\theta}_{i}^{*}$ \citep{lueckmann2021benchmarking}, so we compare Neural Proposal with additional metrics, such as Classifier-based 2-Sample Testing (C2ST) \citep{lueckmann2021benchmarking}, Maximum Mean Discrepancy (MMD) \citep{greenberg2019automatic}, Inception Score (IS) \citep{salimans2016improved}, Simulation-Based Calibration (SBC) \citep{papamakarios2019sequential}, and Median distance (MEDDIST) \citep{papamakarios2019sequential}. C2ST is a classification accuracy and MMD measures the distributional discrepancy between the ground-truth and approximate posteriors. IS measures the diversity of the approximate posterior. SBC qualitatively measures the quality of approximate posterior based on a statistical theory. MEDDIST is the median distance between simulated outcomes from the real-world observation.

\paragraph{Simulation Budget} In the community of \textit{likelihood-free inference}, there is no common consensus on the budget selection. The number of budgets varies by papers and by simulations. For instance, SNL \citep{papamakarios2019sequential} and APT \citep{greenberg2019automatic} select the budget to be $N=1,000$ with $10-40$ rounds for all of their experiments. However, we highlight that such budget is far from optimal on various simulations. On the other hand, AALR \citep{hermans2019likelihood} chooses $N=10,000$ with $100$ rounds of inference, and this requires a total of one million simulation running for the posterior inference, which is highly impractical for expensive simulations. Throughout our experiments, we fix 10 rounds for the inference iterations. For the budget, we fix $N=1,000$ to comply the common practice \citep{papamakarios2019sequential, greenberg2019automatic}, except for M/G/1 with $N=100$.

\begin{figure}[t]
    \centering
    \includegraphics[width=.5\linewidth]{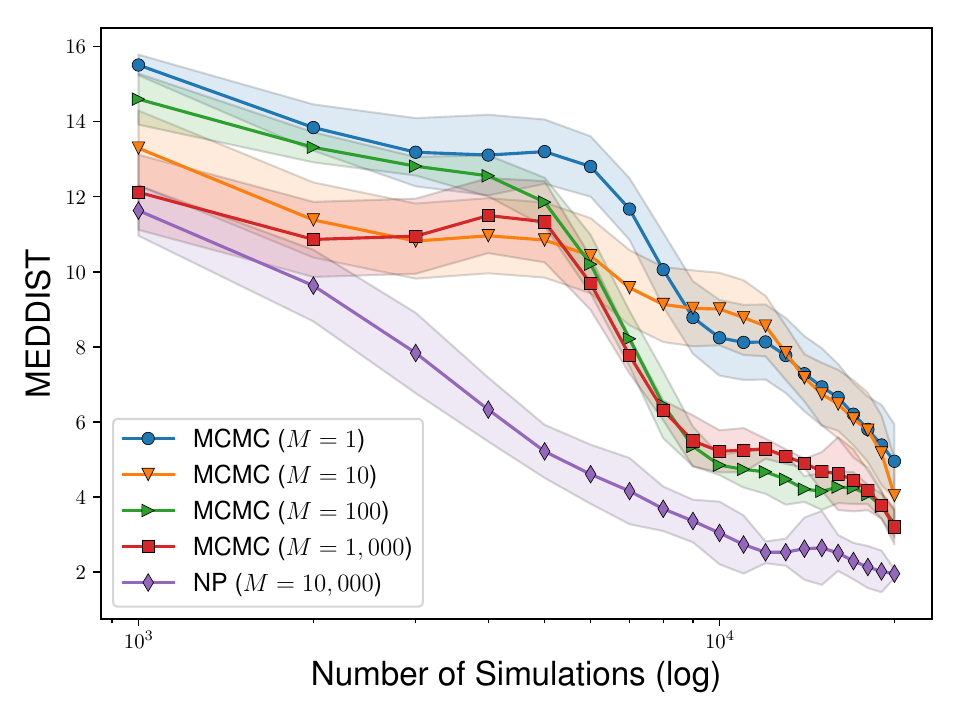}
    \vskip -0.05in
    \caption{MEDDIST on CLV.}
    \label{fig:meddist}
    \vskip -0.1in
\end{figure}

\subsection{Results}

\paragraph{Comparison after Inference} We present the quantitative performances after multi-rounds of inference in Tables \ref{tab:performance} and \ref{tab:performance_real} under 30 replications, and the qualitative SBC in Figure \ref{fig:CLV}. These experiments give three implications.
\begin{itemize}\setlength\itemsep{0.2em}
    \item Neural Proposal combined with SNL/AALR performs the best out of the sampling algorithms for the simulation input, such as MCMC and active learning.
    \item Neural Proposal outperforms MCMC with the maximal number of chains, i.e., $M=N$.
    \item SNL joined with Neural Proposal largely outperforms the posterior modeling approach of APT.
    \end{itemize}

\begin{figure}[t]
\vskip -0.05in
    \centering
    \subfigure[MMD by $\#$ chains]{\includegraphics[width=.495\linewidth]{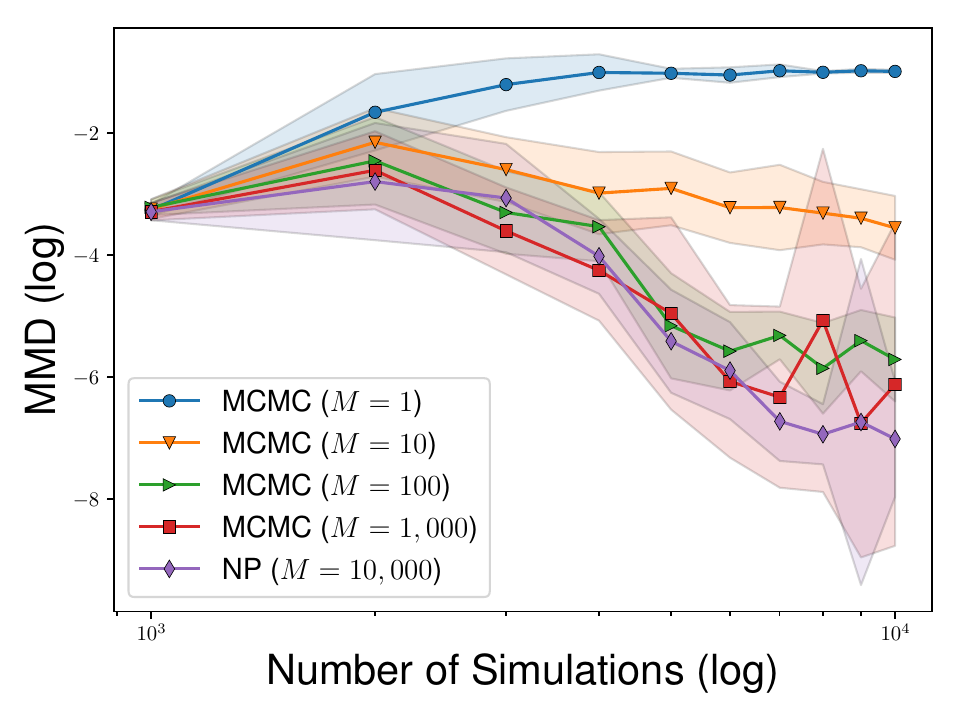}}
    \centering
    \subfigure[Missed Mode by $\#$ modes]{\includegraphics[width=.495\linewidth]{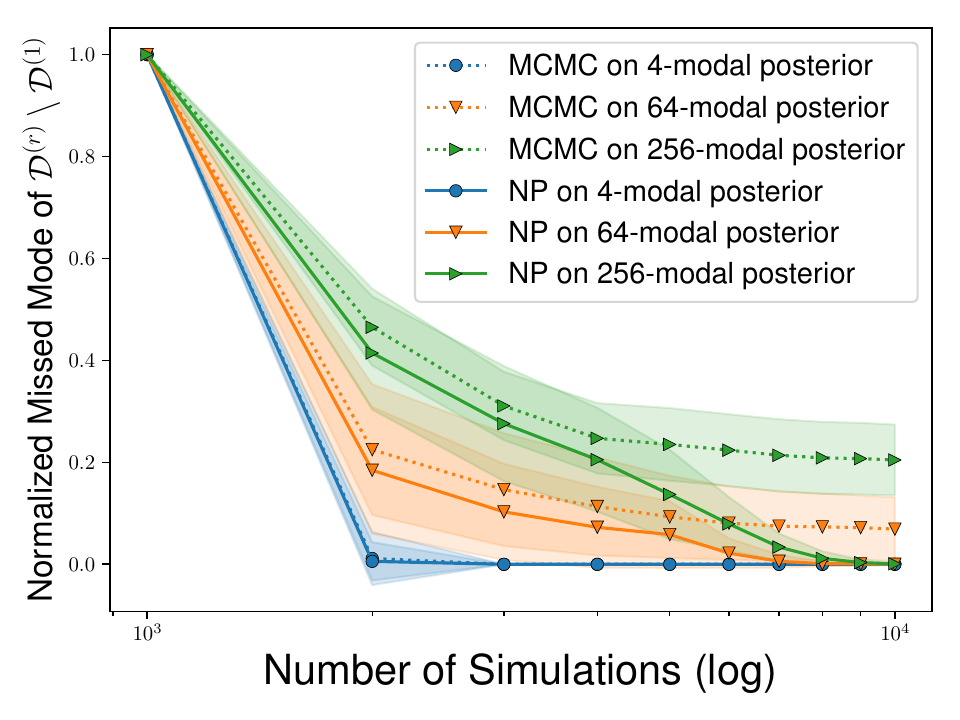}}
    \vskip -0.1in
    \caption{Study on SLCP-16 with (a) $\#$ chains and (b) $\#$ modes inferred by SNL.}
    \label{fig:ablation}
    \vskip -0.1in
\end{figure}

\paragraph{Comparison by Round} Figure \ref{fig:performance_by_round} compares NP with baselines by round. Figure \ref{fig:performance_by_round} implies that SNL/AALR with NP outperforms their counterparts with MCMC. Figure \ref{fig:performance_by_round}-(c,d) indicates that NP significantly mitigates the mode collapse issue. It is worth noting that APT captures the modes the fastest among algorithms including our NP, but SNL with NP eventually outperforms APT in the inference quality. Figure \ref{fig:meddist} compares the median distance of NP against MCMC on CLV simulation.

\begin{figure*}[t]
    \centering
    \subfigure[Groundtruth posterior]{\includegraphics[width=.24\linewidth]{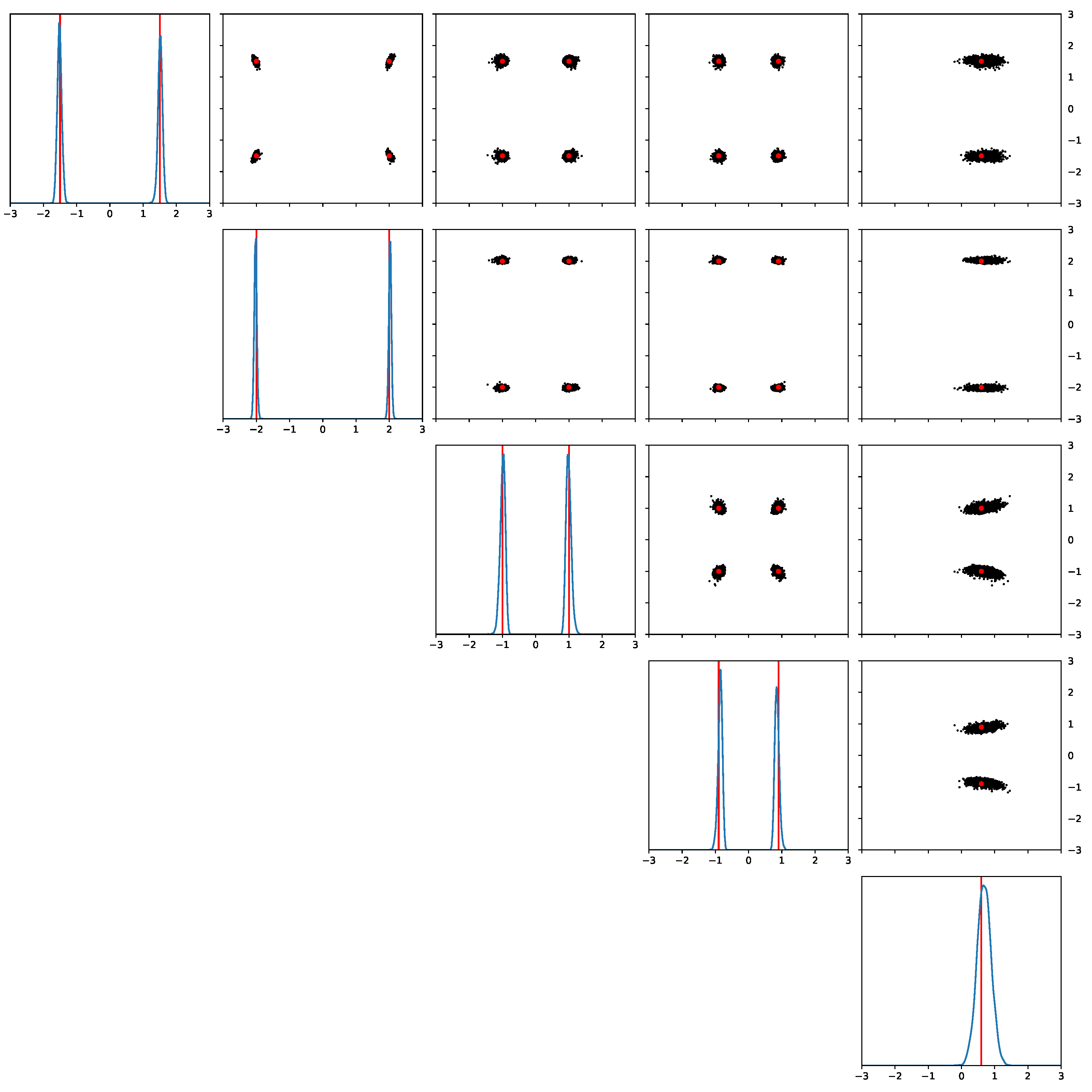}}
    \hspace{1.em}
    \subfigure[Multi-chain MCMC ($M=1,000$)]{\includegraphics[width=.24\linewidth]{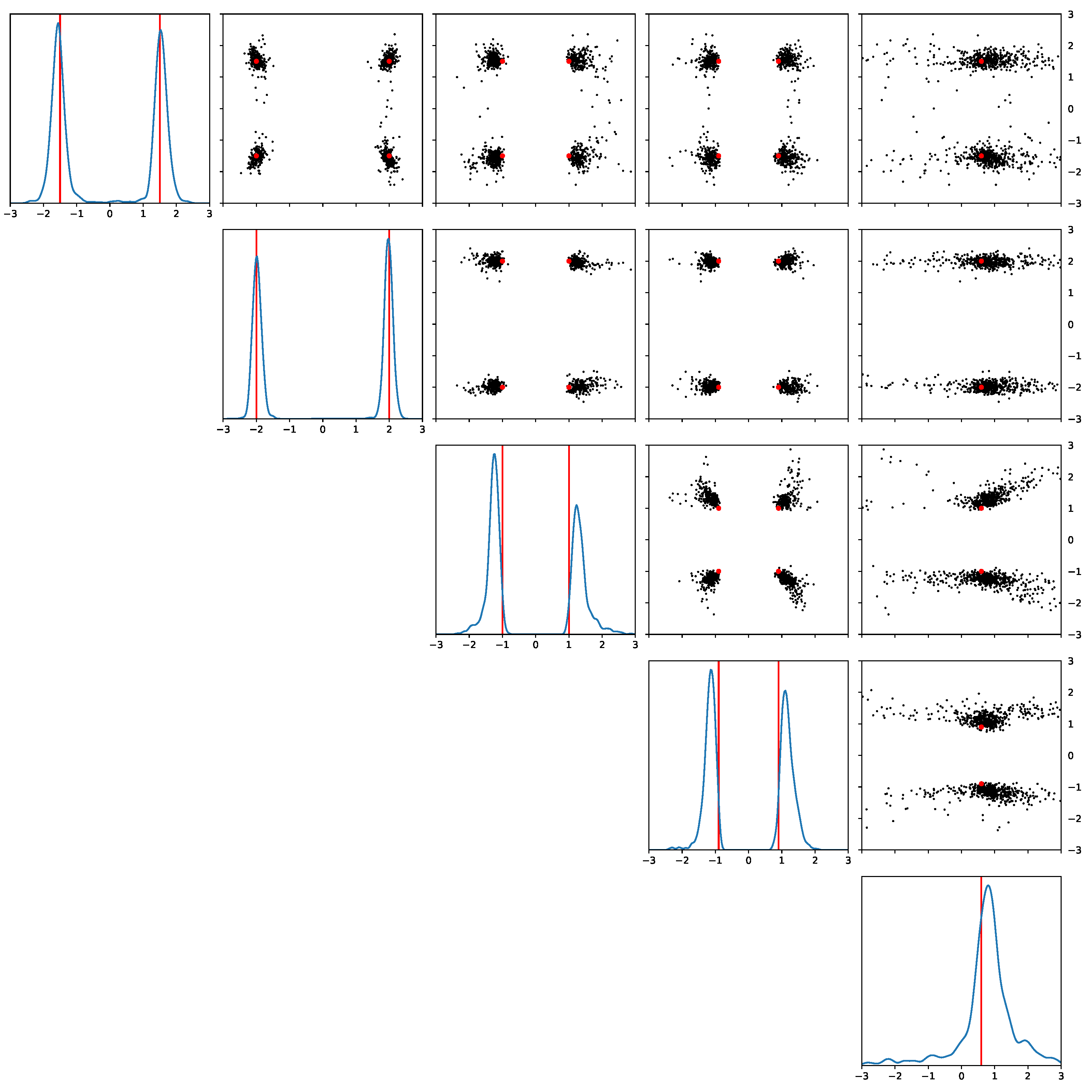}}
    \hspace{1.em}
    \subfigure[Neural Proposal ($M=10,000$)]{\includegraphics[width=.24\linewidth]{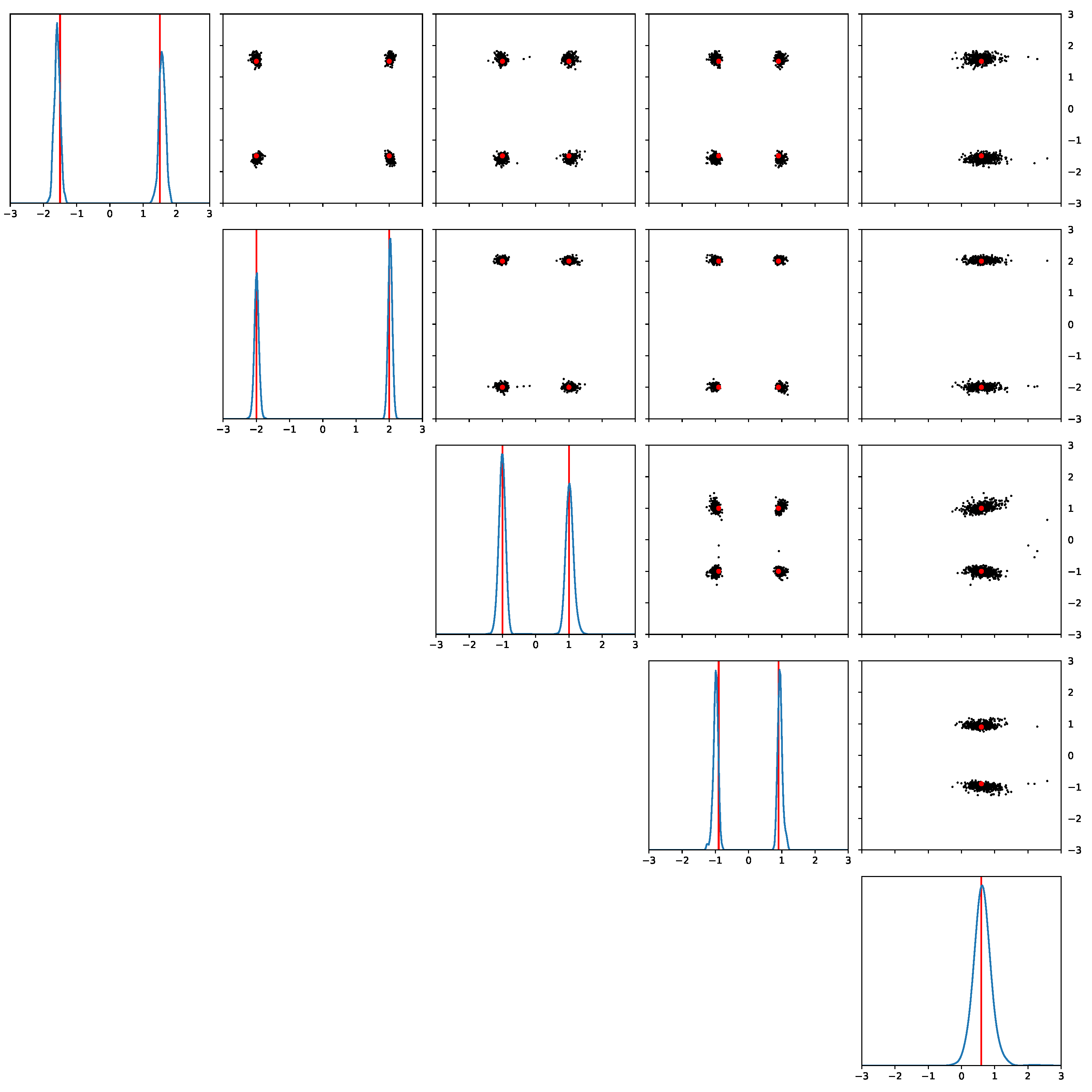}}
    \vskip -0.1in
    \caption{Approximate posteriors inferred by SNL with (b) $1,000$-chained MCMC and (c) NP, and (a) the ground-truth posterior on SLCP-16.}
    \label{fig:approximate_posterior}
    \vskip -0.1in
\end{figure*}

\paragraph{Ablation on Number of Chains} Figure \ref{fig:ablation}-(a) implies that MCMC is improving its performance as $M$ increases up to $M=1,000$, showing the consistency with Theorem \ref{prop:monotone}. From $M\le N=1,000$, MCMC with $M=1,000$-chains becomes the maximally efficient sampler with MCMC. Figure \ref{fig:ablation}-(a) illustrates that NP with $M=10,000$ samples out of $M=10,000$-chains outperforms MCMC. Moreover, Table \ref{tab:performance_chain} presents that SNL with NP consistently improves by $M$.
\begin{table}[t]
    \renewcommand{\arraystretch}{1.4}
    \caption{NLTP inferred by SNL with NP with varying $M$.}
    \label{tab:performance_chain}
    \scriptsize
    \vskip -0.05in
                \resizebox{.99\linewidth}{!}{
                    \begin{threeparttable}
                        \begin{tabular}{ccccccc}
                            \toprule
                            Simulation & M=10 & M=100 & M=1,000 & M=2,500 & M=5,000 & M=10,000\\\midrule
                            \multirow{2}{*}{SLCP-16} & 93.21 & -15.20 & -29.64 & -31.16 & -31.72 & \textbf{-32.11}\\
                            & ($\pm$8.47) & ($\pm$19.39) & ($\pm$30.79) & ($\pm$25.31) & ($\pm$18.43) & ($\pm$19.70)\\\midrule
                            \multirow{2}{*}{SLCP-256} & 37.57 & 30.80 & 18.35 & 17.31 & 13.57 & \textbf{12.30}\\
                            & ($\pm$1.12) & ($\pm$2.81) & ($\pm$8.11) & ($\pm$5.36) & ($\pm$3.09) & ($\pm$2.58)\\
                            \bottomrule
                        \end{tabular}
                    \end{threeparttable}
                }
    \vskip -0.1in
\end{table}

\paragraph{Ablation on Number of Posterior Modes} Figure \ref{fig:ablation}-(b) presents that the missed modes (normalized by the total number of modes) of the approximate posterior inferred by SNL with MCMC/NP samplers. Figure \ref{fig:ablation}-(b) shows the curve of NP is consistently under that of MCMC, regardless of the number of modes. In particular, unlike MCMC which misses some of the modes, NP always finds every modes. 

\paragraph{Samples from Approximate Posterior} Figures \ref{fig:approximate_posterior} and \ref{fig:Example} show that the samples of the approximate posteriors inferred by SNL with MCMC/NP samplers. The samples from the Neural Proposal are distributed the closest to the samples from the ground-truth posterior, compared to MCMC.

\section{Conclusion}

This paper suggests NP that is asymptotically exact to the proposal distribution. NP draws i.i.d. samples via a feed-forward fashion, and this i.i.d. nature helps construct data that is more efficient for posterior inference. In contrast to MCMC, NP is free from the auto-correlation and the mode degeneracy. NP significantly improves the inference quality in every simulations.

\bibliographystyle{elsarticle-num-names}
\bibliography{reference.bib}

\end{document}